\documentclass[twoside,twocolumn,tightenlines, showpacs,aps,prb]{revtex4}
\usepackage{graphicx}
\usepackage[makeroom]{cancel}
\usepackage{float}
\newcommand{\be}{\begin{equation}}
\newcommand{\ee}{\end{equation}}
\begin{document}

\title{Klein Paradox in Chaotic Dirac Billiards}

\author{A. F. M. Rodrigues da Silva$^1$, M. S. M. Barros$^1$, A. J. Nascimento J\'unior$^2$ A. L. R. Barbosa$^3$, J. G. G. S. Ramos$^1$ }

\affiliation{$^1$  Departamento de F\'isica, Universidade Federal da Para\'iba, 58051-970 Jo\~ao Pessoa, Para\'iba, Brazil\\
$^2$ Departamento de Engenharia El\'etrica, Universiade Federal de Pernambuco, 50670-901 Recife, Pernambuco, Brazil\\
$^3$ Departamento de F\'{\i}sica, Universidade Federal Rural de Pernambuco, 52171-900 Recife, Pernambuco, Brazil}

\date{\today}

\begin{abstract}

We investigate the Schr\"odinger (non-relativistic) and the Dirac (``relativistic") billiards in the universal regime. The study is based on a non-ideal quantum resonant scattering numerical simulation. We show universal results that reveal anomalous behavior on the conductance, on the shot-noise power and on the respective eigenvalues distributions. In particular, we demonstrate the Klein's paradox in the graphene and tunable suppression/amplification transitions on the typical observables of the quantum dots.

\end{abstract}


\maketitle

\section{Introduction}

The study of electronic mesoscopic devices is a very attracting subject, deserving the attention of several groups, both theoretical and experimental [\onlinecite{Montambaux,beennaker_rev,Mello_book,Heinzel,Gustavsson_Rev}]. The main aim of the investigations is to understand the quantum electronic transport through observables as the conductance, the universal conductance fluctuations [\onlinecite{fluc1,Montambaux,fluc3}] and the shot-noise power [\onlinecite{noise1,noise2}]. They are affected by tunnel barriers, magnetic fields, confining geometry and so on [\onlinecite{Montambaux,Gustavsson_Rev}]. One of the most studied device is known as ballistic chaotic quantum billiard (QB) [\onlinecite{Heinzel,Ashoori,RBM,Ramos,Dietz,RBH,FSM}]. There are two relevant experimental examples of the QB, the two-dimensional electron gas and the single-layer graphene quantum billiard. The main difference between the two is the wave functions behavior of electrons inside them, the former devices are described by Schr\"odinger equation while the later devices are described by Dirac equation. Therefore, it is appropriate to name them as ballistic Schr\"odinger billiard and ballistic Dirac billiard [\onlinecite{grebogi1,grebogi2,Barros}], respectively.

The random borders of the QB in the regime of strong confinement lead to chaos, such that observables assume aleatory values. In spite of this, the statistics of observables associated with QBs assume universal values, i.e., they depend only on the fundamental symmetries and are independent of the microscopic details of the system. The distributions of the observables lead to remarkable phenomena such as phase transitions associated with Gaussian interactions [\onlinecite{dist1,dist2}], changes in the Fano factor associated with the shot-noise power [\onlinecite{dist3}] and to universality in the conductance eigen-values ​​[\onlinecite{dist4}] and in all the statistics cumulants [\onlinecite{dist5}]. The study of the distributions for conductance in Dirac billiards is still a subject of intense investigations. In part, it is assumed that a large number of open channels in the leads  (semi-classical limit) that connect the QD with the reservoirs leads to the equivalence between the observables of the Dirac billiards and the Schr\"odinger billiards ones.

The question remains about the role of tunnelling (non-ideal contacts) in the Dirac billiards and the emergence of phenomena that provide strong quantum signals in graphene also with a wide lead. On the other hand, the shot-noise power (noise at zero temperature) is an observable that carries direct information about the tunneling process associated to the discretisation of the electronic current. Therefore, a detailed study of both the conductance and the noise of Dirac billiards in the universal regime will clearly show the peculiarities of graphene/chirality compared to other materials in condensed matter physics. The obtention of theoretical results in the presence of a barrier (arbitrary tunnelling rate) is a hard task [\onlinecite{richter1,richter2,jacquod}]. The analytical techniques usually involve expansion in the semi-classical limit, that is, a high number of scattering channels. However, as the number of open channels increases, the chirality of the graphene is lost and it merges with a usual Schr\"odinger billiard [\onlinecite{ramos1}]. Therefore, numerical investigations can provide important signals of chirality in a regime for which the barriers are arbitrary and the number of channels is very low [\onlinecite{dietz}].

In the random matrix theory (RMT) framework, the universal scattering matrix which describes the electronic transport through the Schr\"odinger billiard is a member of Wigner-Dyson ensembles while the correspondent one associated with the Dirac billiard is a member of the Chiral ensemble [\onlinecite{AZ,JB,Metha,Ver}]. The RMT approach is insensitive to the irrelevant microscopic details of the system, but is strongly affected by the intrinsic symmetries of the corresponding Hamiltonian, such as time-reversal symmetry, spin-rotation symmetry, particle-hole symmetry and sub-lattices/chiral symmetry. Regarding the Wigner-Dyson ensembles, it is divided in three ensembles: orthogonal ensemble (OE) which preserves time-reversal symmetry and spin-rotation symmetry ($\beta=1$); unitary ensemble (UE) which has time-reversal symmetry broken by magnetic field ($\beta=2$); symplectic ensemble (SE) which preserves time-reversal symmetry and has spin-rotation symmetry broken ($\beta=4$). The Chiral ensembles is also divided in tree ensembles: chiral orthogonal ensemble which preserves time-reversal symmetry, spin-rotation symmetry, particle-hole symmetry and sub-lattices/chiral symmetry ($\beta=1$); chiral unitary ensemble which has time-reversal symmetry and particle-hole symmetry broken and preserves sub-lattices/chiral symmetry ($\beta=2$); chiral symplectic ensemble which preserves time-reversal symmetry, particle-hole symmetry and sub-lattices/chiral symmetry but has spin-rotation symmetry broken ($\beta=4$).

The seminal papers by Klein, Souter, and Hund investigated scattering due to high potentials compared to the incident energy [\onlinecite{klein1}]. They concluded that, unlike the Schr\"odinger equation, the scattering provoked by the Dirac equation may lead to an amplification of the  probability amplitude of the tunneled relativistic electron. This paradox was initially solved considering that the very energetic electron spontaneously creates positrons that reemit electrons in order to maintain the constant current. In condensed matter physics, the collective effects [\onlinecite{klein2,klein3}] can lead, as we shall show, to the Klein's paradox in the presence of chaos. Therefore, the aim of our investigation is to show the emergence of Klein paradox in chaotic Dirac billiards.

\begin{figure}
\centering
\includegraphics[width=0.4\textwidth]{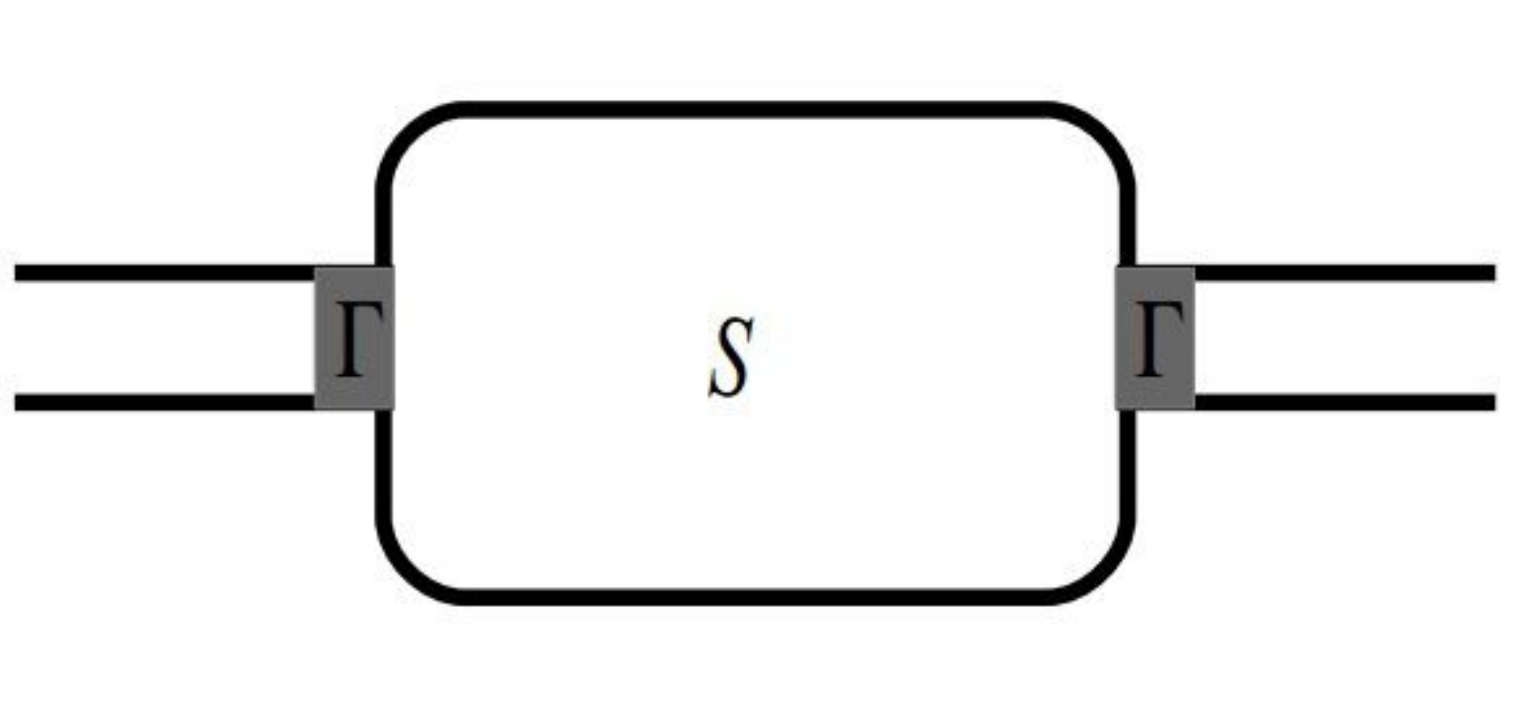}
\caption{The chaotic quantum billiard connected to two leads by tunnel barriers ($\Gamma$). The electronic transport is described by a scattering matrix, $\mathcal{S}$, for both the Schr\"odinger billiard (without sub-lattices/chiral symmetry) and the Dirac billiard (with sub-lattices/chiral symmetry).} \label{Fig0}
\end{figure}

In this work, we perform a complete numerical study of the electronic transport statistics for both Schr\"odinger billiard and Dirac billiard connected to non-ideal leads by tunnel barriers, as depicted in the Fig.(\ref{Fig0}). We investigate the role of the tunnel barriers on the statistic of conductance (and its fluctuations) and the shot-noise power. We dedicate special attention to the regime in which the leads have few number of open channels ($N=1,2,3$), and we observe a very distinct quantum statistics behaviour comparing the Schr\"odinger billiard with the Dirac billiard. We also show that the regime of few open channels in the presence of tunnel barriers is strongly affected by quantum interference effects that can be measured experimentally. Furthermore, the study brings information concerning electronic transport which were never obtained by analytical methods.

The work is organised as in the following: In Section II we introduce the numerical scattering approach and write out the probability distributions of Hamiltonian elements in the framework of RMT; In Section III we write a brief review for pedagogical reasons of Schr\"odinger billiard and Dirac billiard without tunnel barriers and compare our numerical results with known analytical results; In the Sections IV and V we analyse the effects of tunnel barriers over the statistic  of conductance and shot-noise for Schr\"odinger billiard and Dirac billiard, respectively and, finally, we show an interesting effect caused by tunnel barriers over conductance of Dirac billiard; In Section VI, we conclude.

\section{Scattering Approach}

In this section, we introduce the scattering approach applied to the study of the electronic transport through a quantum chaotic billiard connected to leads by non-ideal contacts, as depicted in the Fig.(\ref{Fig0}). Furthermore, we study the effects of sub-lattices/chiral symmetry over electronic transport in the device. 

The standard form of the scattering matrix is a composition of the reflection ($r$) and the transmission ($t$) probabilities amplitudes blocks determined by the channels of the leads
\begin{equation}
\mathcal{S}=\left[\begin{array}{cc}
    r &  t \\ 
     t'& r' \\ 
            \end{array}\right],
\end{equation}
in which $r$, $t$, $t'$ and $r'$ have the dimensions $N_1 \times N_1$, $N_1 \times N_2$, $N_2 \times N_1$ and $N_2 \times N_1$, respectively. The number of open channels is a function of the width, $a_i$, of the leads, $N_i \propto a_i$. Accordingly, the scattering matrix has dimension $N\times N$ and it can be obtained as a function of the Hamiltonian as in the following [\onlinecite{MW}]
\begin{equation}
\mathcal{S}(\epsilon)= \textbf{{1}}-2\pi i\mathcal{W}^{\dagger}(\epsilon-\mathcal{H}+i\pi \mathcal{WW}^{\dagger})^{-1}\mathcal{W}. \label{MatrisS}
\end{equation}\\
The Hamiltonian $\mathcal{H}$ of the target (QD) supports $M$ resonant modes and incorporates the chaotic behavior of the system through its random aspects. Its dimension is $M\times M$ and the Fermi energy of the electron is denoted by $\epsilon$. The deterministic matrix $\mathcal{W}$ describes the contacts (coupling) between the chaotic billiard (resonances) and the leads (channels) and, accordingly, it has dimension $N \times M$. The $\mathcal{W}$ matrix can be divided into two blocks, $\mathcal{W}_1$ with dimension $N_1 \times M$ and $\mathcal{W}_2$ with dimension $N_2 \times M$ in such a way that $\mathcal{W}=(\mathcal{W}_1 \; \mathcal{W}_2)$. The direct processes (prompt transport, that is, transmission processes without the transition through the resonant QD) can be avoided through the imposition of orthogonality condition $$\mathcal{W}_{\alpha} \mathcal{W}_{\beta}= \gamma_{\alpha} \frac{N \Delta}{\pi^2} \delta_{\alpha,\beta},$$ to which $\Delta$ is the mean energy level spacing of the QD and $\gamma_{\alpha}$ is a diagonal matrix written as $\gamma_\alpha= \textrm{diag}(\gamma_{\alpha,1},\dots,\gamma_{\alpha,N})$. The $\gamma_{\alpha}$ matrix can be related with the transmission probabilities $\Gamma_{\alpha,a} \in [0,1]$ of the $a$ channel supported by de lead $\alpha$ throght the relation $\Gamma_{\alpha,a}= \textrm{sech}^2 \left[- \ln \left( \gamma_{\alpha a} \right)/2 \right]$.

For our investigation, we take equals tunnelling probabilities, i.e, $\Gamma=\Gamma_{\alpha,a}, \; \forall (\alpha,a)$. The tunnelling probabilities are also called barriers and are called ideal barriers whenever $\Gamma=1$ (total tunnelling chance) and opaques if $\Gamma=0$ (very remote tunnelling chance). The tunnelling barriers can be implemented in condense matter physics as a result of imperfect contacts between the leads and the QD or controlled using tension gates.

We start studying a chaotic Sch\"ordinger billiard. In the framework of RMT, the Schr\"odinger billiard Hamiltonian is a member of a Wigner-Dyson ensemble with a Gaussian probability distribution given by [\onlinecite{Metha}]
\begin{equation}
P(\mathcal{H})\approx \exp\left[-\frac{\beta M}{4\lambda^2} \textbf{Tr}(\mathcal{H}^2)\right],
\end{equation}
in which the index $\beta={1,2,4}$ is to account the Gaussian Orthogonal Ensemble (GOE), Gaussian Unitary Ensemble (GUE) and Gaussian Symplectic Ensemble (GSE), respectively.

Specifically for the chaotic Dirac billiard, the Hamiltonian supports the sub-lattices/chiral symmetry, i.e., it must satisfy the following anti-commutation relation [\onlinecite{JB,Ver}]
\begin{equation}
\mathcal{H}=-\sigma_{z} \mathcal{H} \sigma_{z},\ \ \ \sigma_{z}=\left[
                      \begin{array}{cc}
                         \bf{1_{M}} & 0  \\
                        0 & \bf{-1_{M}}  \\
                      \end{array}
                    \right]. \label{Hsig}
\end{equation}
which implies that the Hamiltonian is an anti-diagonal matrix
\begin{equation}
\mathcal{H}=\left[
                      \begin{array}{cc}
                        0 & \mathcal{T}  \\
                        \mathcal{T}^{\dag} & 0  \\
                      \end{array}
                    \right]. 
\end{equation}
Hence, the Hamiltonian of the Dirac billiard is a member of the Chiral ensembles and the $\mathcal{T}$-matrix admits the probability distribution given by 
\begin{equation}
P(\mathcal{T})\approx \exp\left[-\frac{\beta M}{4\lambda^2} \textbf{Tr}(\mathcal{T}^2)\right],
\end{equation}
in which the index $\beta={1,2,4}$ is to account the chiral GOE (chGOE), the chiral GUE (chGUE) and the chiral GSE (chGSE), respectively, to account the additional symmetries.

\section{Chaotic Billiard with Ideal Leads} 

For pedagogical reasons, and in order to present a self-contained argumentation, we first explore the  conventional chaotic billiards with ideal contacts. We numerically study the conductance statistics using Landauer-B\"uttiker formulation
\begin{equation}
G= G_0 \textbf{Tr}(tt^{\dag})\label{G}
\end{equation}
in which $G_0=2e^2/h$. The shot-noise power is the amplitude of the current temporal correlation and characterizes the discreteness of the charge transport process. At zero temperature, the shot-noise power is given by [\onlinecite{noise1,noise2}]
\begin{equation}
P= P_0 \textbf{Tr}\left[tt^{\dag}(1-tt^{\dag})\right]\label{P}
\end{equation}
with $P_0=4e^2V/h$.

\begin{figure}[H]
\centering
\includegraphics[width=0.45\textwidth]{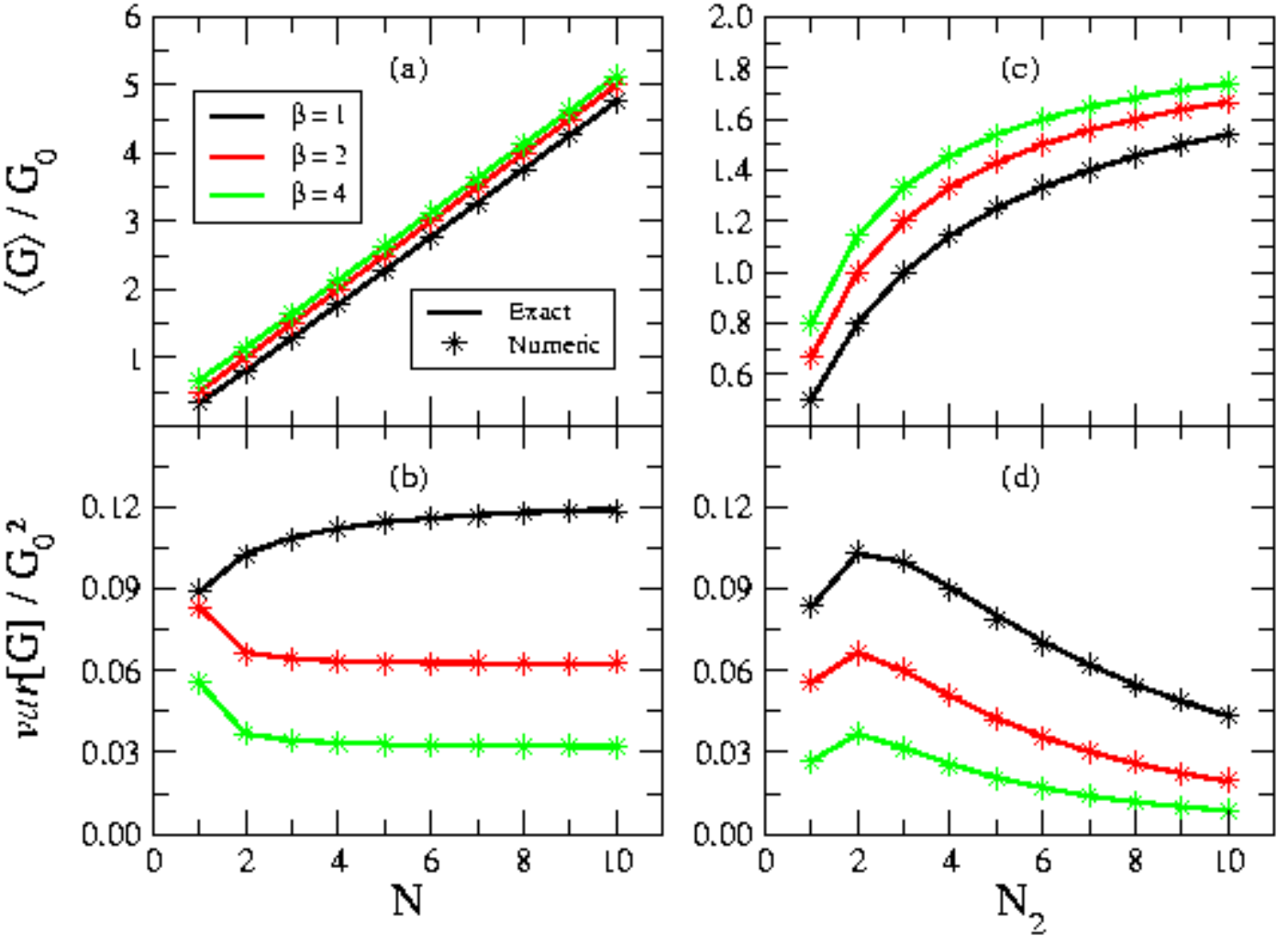}
\caption{The average and variance of conductance for a chaotic Schr\"odinger billiard. The curves (a,b) show the special case of symmetric leads $N_{1}=N_{2}=N$, and (c,d) the case of asymmetric leads  with $N_{1}=2$. The symbols are the numerical results data while the lines are the analytical results of the Eqs.(\ref{GWD}) and (\ref{VGWD}). Notice that the average conductance has a linear behaviour with the width (number of channels) of the leads as the number of channels increase, i.e., an Ohmic behaviour. Nevertheless, the chaotic behaviour still remains as the universal fluctuations indicates.} \label{Fig2pt1}
\end{figure}

Using the previously described scattering approach, we perform a numerical simulation using an ensemble of ${\cal N}=10^8$ Hamiltonians with $400$ resonances and, together with the Eq.(\ref{G}), we plot the average of conductance and its variance for Schr\"odinger billiard with symmetric ($N_1=N_2=N$) and asymmetric ($N_1=2$) leads, as depicted in the Fig.(\ref{Fig2pt1}). The analytical results are well known and given by [\onlinecite{Mello_book,Brouwer}]
\begin{equation}
\langle G\rangle=G_0\frac{ N_1N_2}{ N_T - 1 + 2/\beta}\label{GWD}
\end{equation}
and 
\begin{equation}
\frac{var[G]}{G_0^2}=\textstyle \frac{2N_{1}N_{2}(N_{1}-1+2/\beta)(N_{2}-1+2/\beta)}{\beta(N_{T}-1+4/\beta)(N_{T}-2+2/\beta)(N_{T}-1+2/\beta)^2}\label{VGWD}
\end{equation}
as a function of the index $\beta=1,2,4$. The analytical and the numerical results are in agreement.

\begin{figure}[H]
\includegraphics[width=0.45\textwidth]{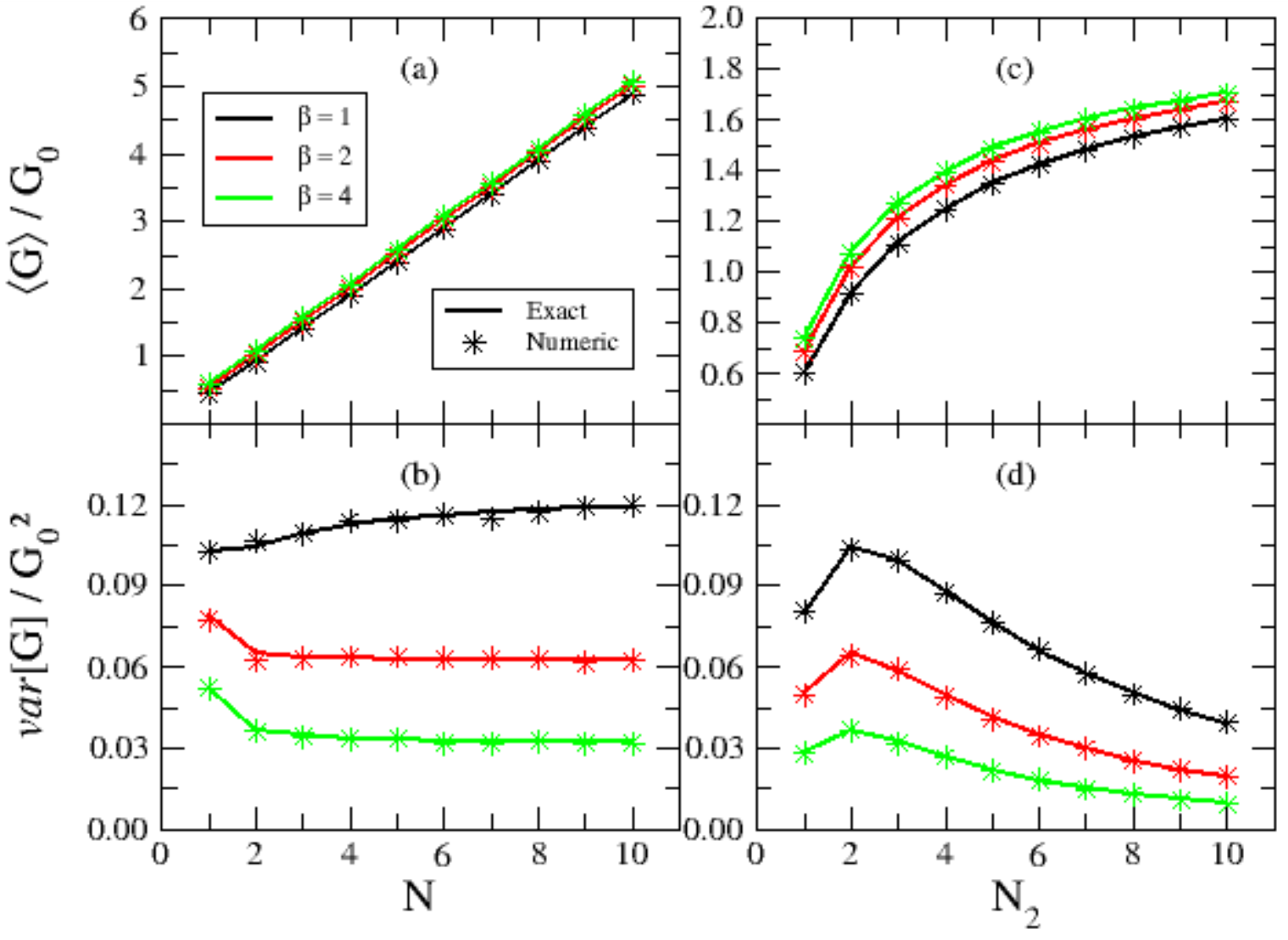}
\caption{The average and variance of conductance for a chaotic Dirac billiard. The plots (a) and (b) describes symmetric leads $N_{1}=N_{2}=N$, and (c) and (d) asymmetric ones with $N_{1}=2$. The symbols are numerical calculation data while the lines are analytical results of Eqs.(\ref{GCH}) and (\ref{VGCH}).} 
\label{Fig3pt1}
\end{figure}

Furthermore, in the Fig.(\ref{Fig3pt1}), we plot the average of conductance and its variance for the Dirac billiard with symmetric ($N_1=N_2=N$) and asymmetric ($N_1=2$) leads to direct comparison with Schr\"odinger billiard Fig.(\ref{Fig2pt1}). The analytical results are also known and given by [\onlinecite{Barros}] Eq.(\ref{GCH}, \ref{VGCH}), which is in agreement with the numerical results (see Fig.(\ref{Fig3pt1})).

\begin{equation}
\langle G\rangle=G_0\frac{4\beta N_1N_2N_T}{(\beta N_T+1)(2N_T-1)}
\label{GCH}
\end{equation}

and

\rule{0.0pt}{0pt}
\begin{equation}
\frac{var[G]}{4e^2/h^2}= \left\{\begin{array}{lll}
			   &\beta_{1},\ \ \beta=1 \\
                                   &\beta_{2},\ \ \beta=2 \\
			   &\beta_{4},\ \ \beta=4 \\
			   \end{array}\right.  
\label{VGCH}
\end{equation}

\noindent in which,\\

\begin{equation}
\textstyle \beta_{1}=\frac{16N_{1}N_{2}N_{T}(3 + 2N_{T}^3+ 4N_{1}N_{2}N_{T}^2 - 4N_{T} - 4N_{1}N_{2} - 5N_{1}^2 - 5N_{2}^2)}{(2N_{T} - 3)(2N_{T} - 1)^2(N_{T} + 3)(N_{T} + 1)^2(2N_{T}+1)};
\end{equation}

\begin{equation}
\textstyle \beta_{2}=\frac{8N_{1}N_{2}(3 + 16N_{1}N_{2}N_{T}^2 - 6N_{T}^2 - 6N_{1}^2 - 6N_{2}^2)}{(2N_{T} - 3)(2N_{T} + 3)(2N_{T} + 1)^2(2N_{T} - 1)^2};
\end{equation}

\begin{equation}
\textstyle \beta_{4}=\frac{32N_{1}N_{2}N_{T}(3 - 16N_{1}N_{2} + 8N_{T} - 20N_{1}^2 - 20N_{2}^2 - 16N_{T}^3 + 64N_{1}N_{2}N_{T}^2)}{(4N_{T} + 3)(4N_{T} + 1)^2(2N_{T} - 3)(2N_{T}-1)^2(4N_{T} - 1)}.
\end{equation}

%\rule{0.0pt}{0pt}
%\begin{equation}
%\scriptstyle \frac{var[G]}{4e^2/h^2}= \left\{\begin{array}{lll}
%			   &\scriptscriptstyle \frac{16N_{1}N_{2}N_{T}(3 + 2N_{T}^3+ 4N_{1}N_{2}N_{T}^2 - 4N_{T} - 4N_{1}N_{2} - 5N_{1}^2 - 5N_{2}^2)}{(2N_{T} - 3)(2N_{T} - 1)^2(N_{T} + 3)(N_{T} + 1)^2(2N_{T}+1)}, \ \ \ \ \ \ \ \ \ \text{$\beta=1$}\\ \\
%                                   &\scriptscriptstyle \frac{8N_{1}N_{2}(3 + 16N_{1}N_{2}N_{T}^2 - 6N_{T}^2 - 6N_{1}^2 - 6N_{2}^2)}{(2N_{T} - 3)(2N_{T} + 3)(2N_{T} + 1)^2(2N_{T} - 1)^2}, \ \ \ \ \ \ \ \ \ \ \ \ \ \ \ \ \ \ \ \ \ \ \ \ \ \ \ \ \text{$\beta=2$}\\ \\
%			   &\scriptscriptstyle \frac{32N_{1}N_{2}N_{T}(3 - 16N_{1}N_{2} + 8N_{T} - 20N_{1}^2 - 20N_{2}^2 - 16N_{T}^3 + 64N_{1}N_{2}N_{T}^2)}{(4N_{T} + 3)(4N_{T} + 1)^2(2N_{T} - 3)(2N_{T}-1)^2(4N_{T} - 1)}, \ \ \ \text{$\beta=4$}\\
%			   \end{array}\right.  
%\label{VGCH}
%\end{equation}

For ideal contacts, notice that the Dirac billiard conductance behaviour is qualitatively similar to the Schr\"odinger billiard ones. In particular, the conductance values $\left< G \right>_{\beta=4}>\left< G \right>_{\beta=2}>\left< G \right>_{\beta=1}$ for all values of $N_i$. The averaged conductance in $\beta=2$ (quantum dot in the presence of a perpendicular magnetic field) has no correction of quantum interference as a consequence of the time-reversal symmetry absence. Therefore, according to the Figs.(\ref{Fig2pt1},\ref{Fig3pt1}) of the ideal regimes, the limit of few open channels preserves the anti-localization as effect of the spin-orbit coupling and the localization in the presence of a magnetic field perpendicular to the sample, in both Schr\"odinger billiard and Dirac billiard.

\begin{figure}[H]
\includegraphics[width=0.45\textwidth]{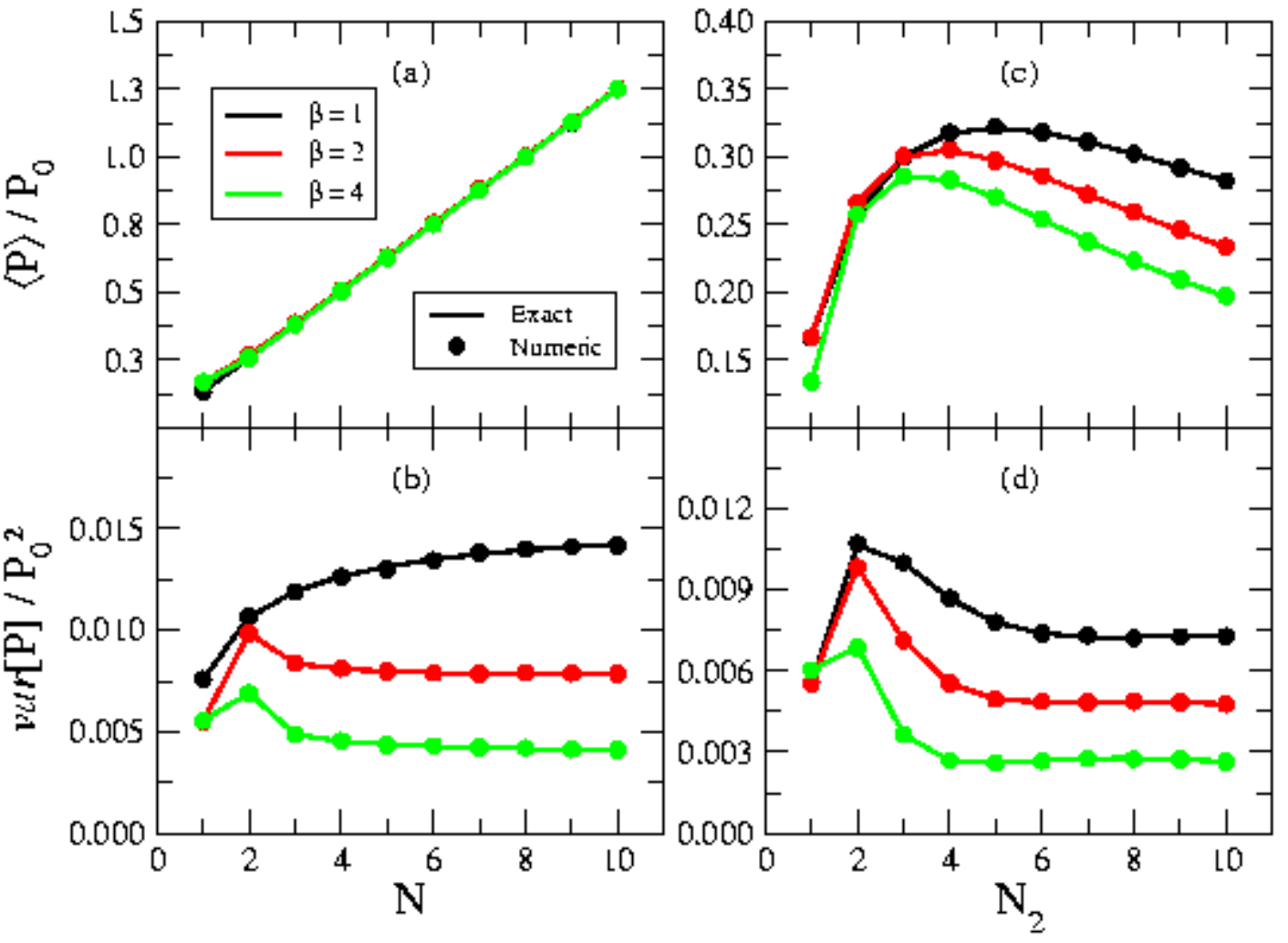}
\caption{The average and variance of shot-noise power for a chaotic Schr\"odinger billiard. The plots (a) and (b) describes symmetric leads $N_{1}=N_{2}=N$. The symbols are a result of the numerical simulation data while the lines are analytical results.} 
\label{Fig4pt1}
\end{figure}

\begin{figure}[H]
\includegraphics[width=0.45\textwidth]{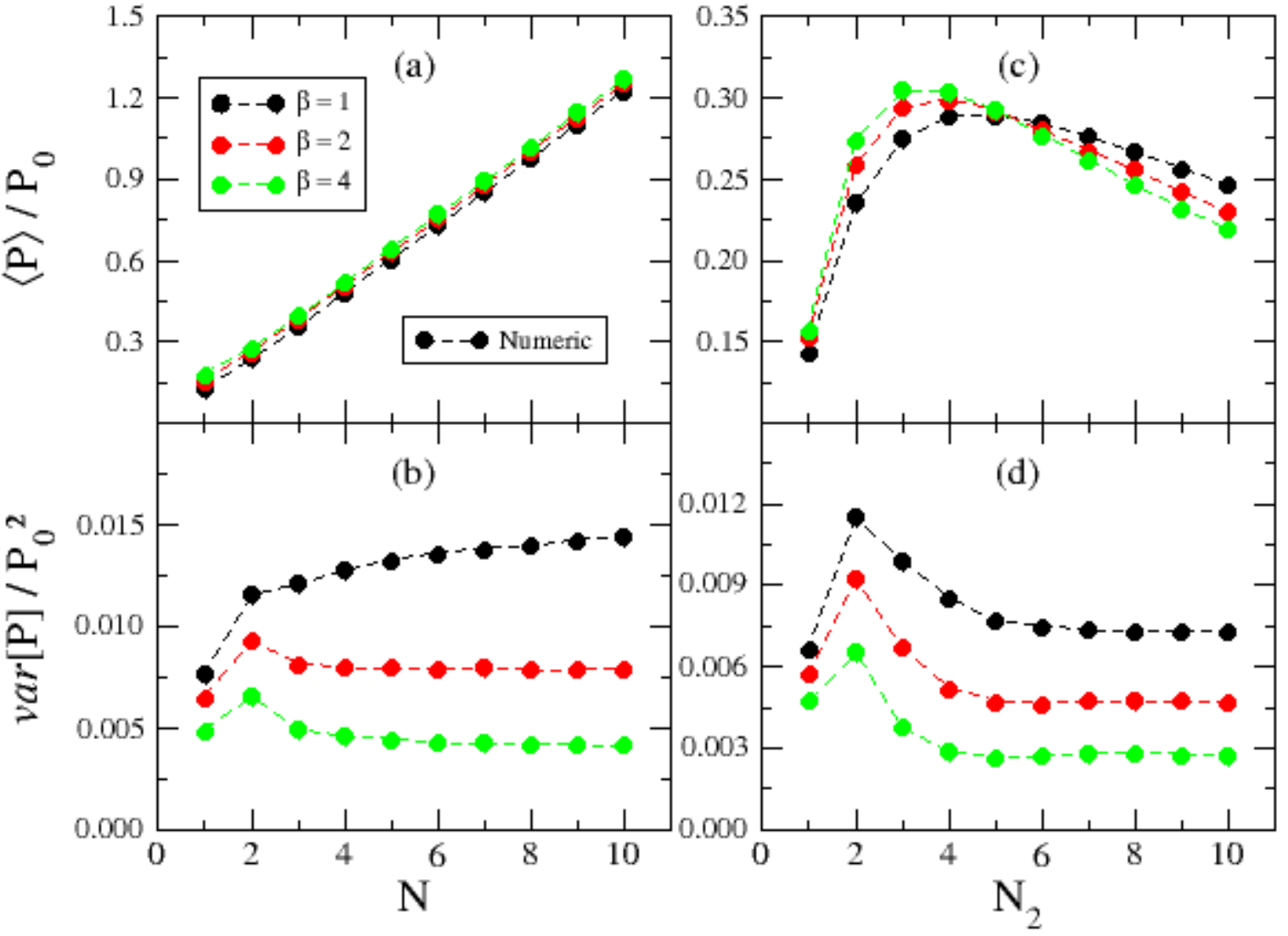}
\caption{The average and variance of shot-noise power for a chaotic Dirac billiard. The plots (a) and (b) describes symmetric leads $N_{1}=N_{2}=N$, and (c) and (d) asymmetric leads with $N_{1}=2$. The symbols are numerical calculation data.} \label{Fig5pt1}
\end{figure}

Using the same numerical scattering approach, we plot also the numerical results for the shot-noise power and its variance for both the Schr\"odinger billiard and the Dirac billiard, as depicted in the Figs.(\ref{Fig4pt1}) and (\ref{Fig5pt1}), respectively.  

Although we are presenting firstly the ideal setup, an intriguing result appears: notice that, as depicted in the Fig.(\ref{Fig5pt1}), the Dirac billiard noise behaviour has a clear distance when comparing the different ensembles, which does not occur in the Schr\"odinger billiard. Therefore, the quantum backscattering mechanisms are more compelling in materials such as graphene. In addition, note a clear inversion of the noise values comparing the three ensembles: according to Fig.(\ref{Fig5pt1}.c), up to $N = 5$, quantum interference amplifies the amplitude in $\beta=4$ for the Dirac billiard while for the Schr\"odinger billiard it suppresses the amplitude as depicted in the Fig.(\ref{Fig4pt1}.c). Therefore, quantum interference occurs inversely for the Dirac billiard shot noise power in the regime of few open channels.

\section{Non-Ideal Schr\"odinger Billiards} 

In this section, we show how non-ideal contacts strongly affects the behaviour of quantum transport through a chaotic Schr\"odinger billiard.

\subsection{Conductance and Shot-Noise }

The analytical results to the average of both conductance and the shot-noise power are known only in the semi-classical regime ($N\gg 1$). 

\begin{figure}[H]
\includegraphics[width=0.45\textwidth]{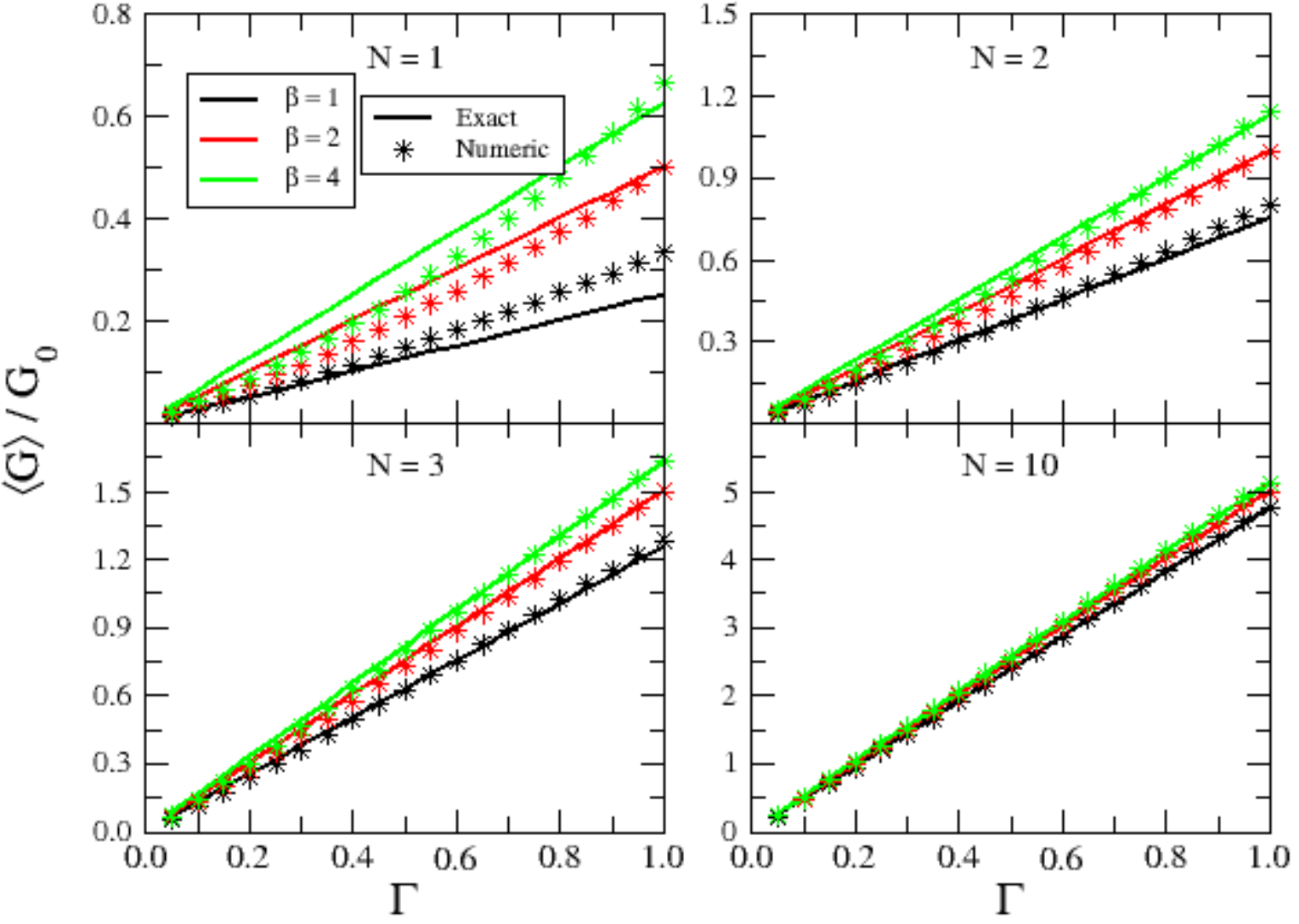}
\caption{The average of conductance as a function of tunnel barrier $\Gamma$ for Schr\"odinger billiard with symmetric leads $N=1,2,3$ and 10. The symbols are numerical calculation data while the lines are analytical results of Eq. (\ref{GT}).} \label{Fig7pt1}
\end{figure}

\begin{figure}[H]
\includegraphics[width=0.45\textwidth]{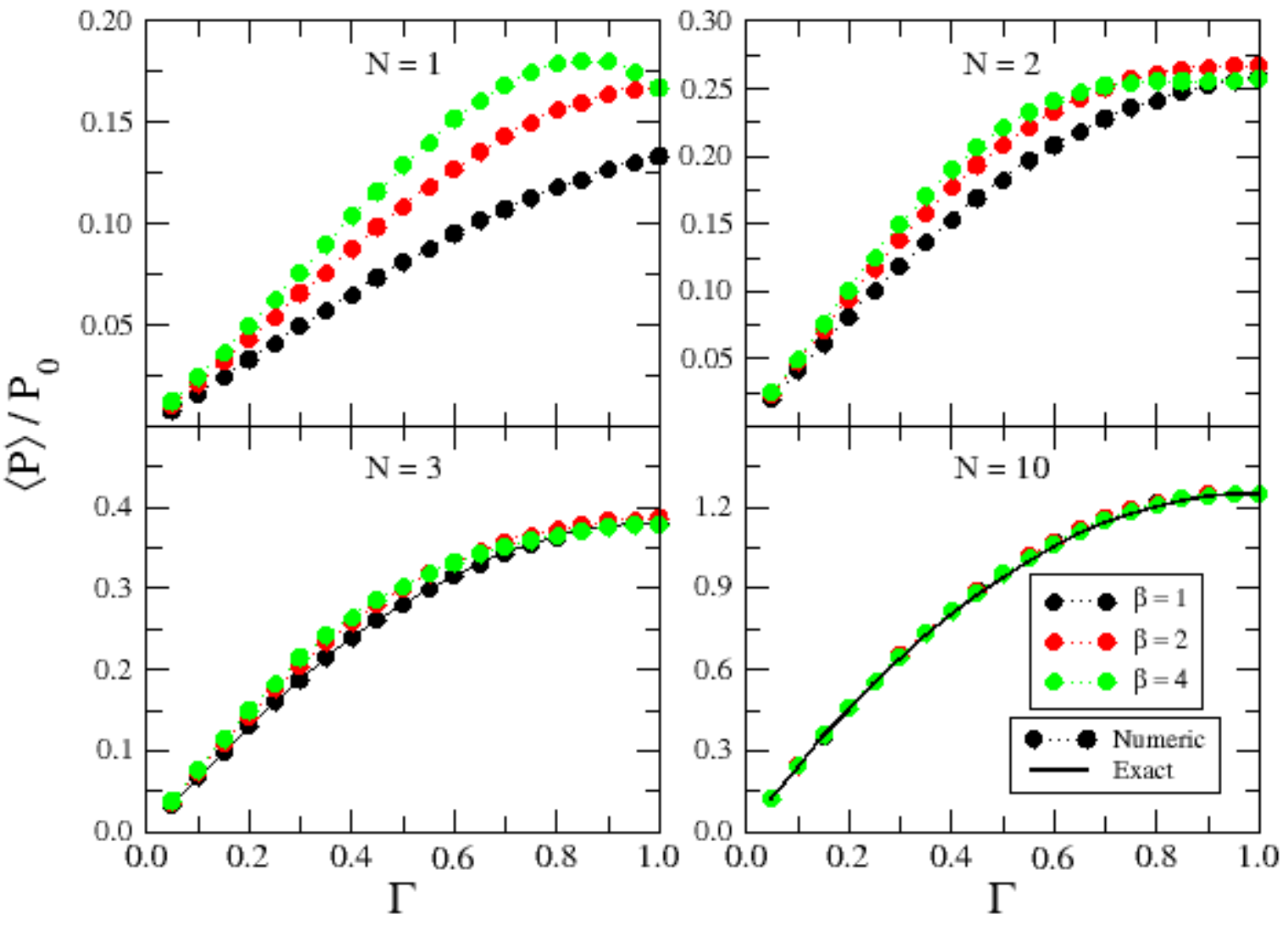}
\caption{The average of shot-noise power as a function of tunnel barrier $\Gamma$ for Schr\"odinger billiard with symmetric leads $N=1,2,3$ and $10$. The symbols are numerical calculation data while the lines are analytical results of Eq. (\ref{PT}). } \label{Fig8pt1}
\end{figure}

\begin{figure}[H]
\includegraphics[width=0.45\textwidth]{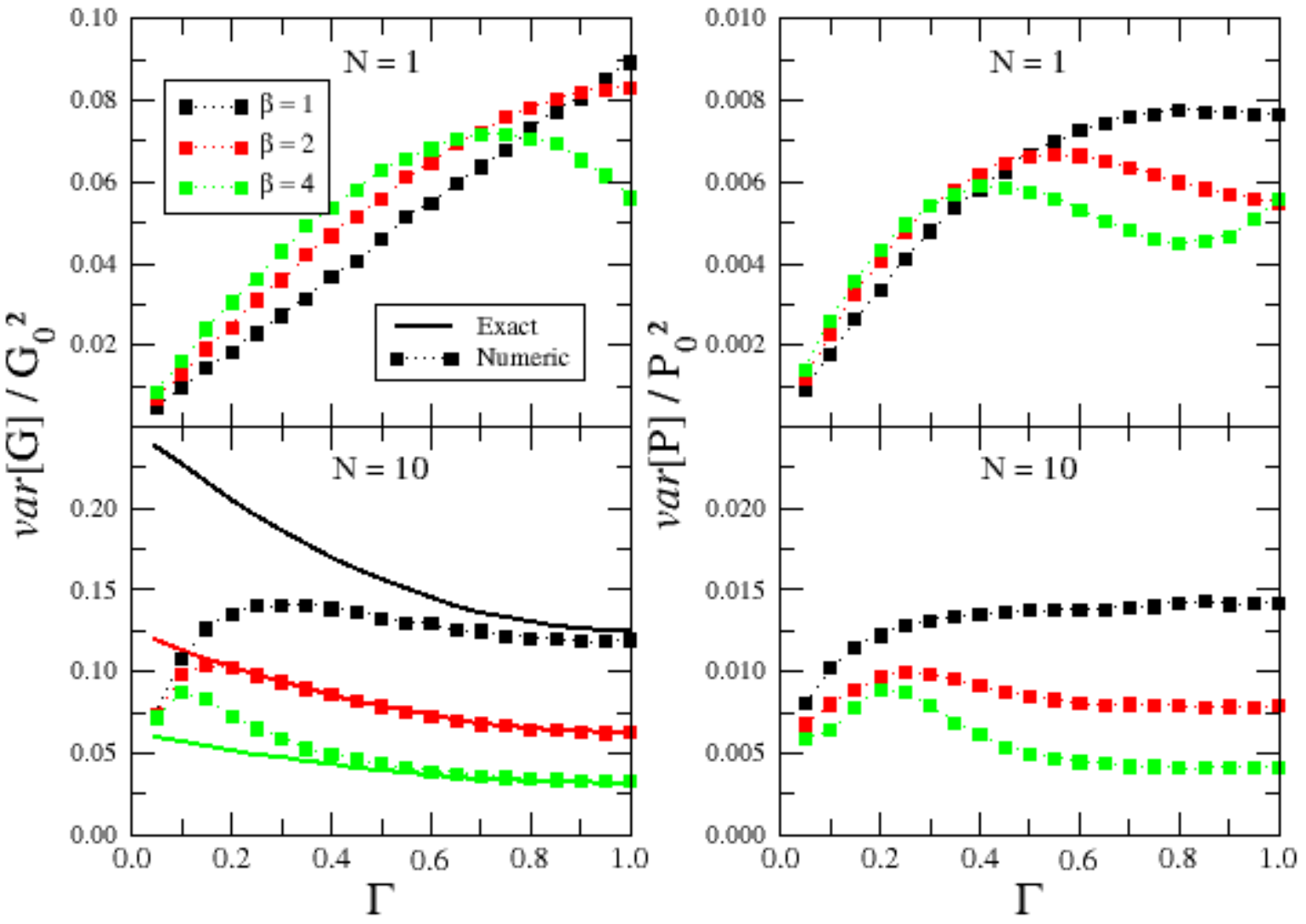}
\caption{The variance of conductance and shot-noise power for chaotic Schr\"odinger billiard.} \label{Fig9pt1}
\end{figure}

\noindent For symmetric leads, the results are given by [\onlinecite{Brouwer}]
\begin{eqnarray}
\frac{\langle G\rangle}{G_0}&=&\frac{N\Gamma}{2}+\frac{\Gamma}{4}\left(1-\frac{2}{\beta}\right) \label{GT}
\end{eqnarray}
and [\onlinecite{RBM}]
\begin{eqnarray}
\frac{\langle P\rangle}{P_0}&=&\frac{N\Gamma}{8}\left(2-\Gamma\right) \label{PT}
\end{eqnarray}
in which $\Gamma \in [0,1]$ is the tunnel barrier. In the Figs.(\ref{Fig7pt1}) and (\ref{Fig8pt1}), we plot our numerical simulation results for the average of conductance and the shot-noise power as a function of the tunnel barrier. The analytical and numerical results are in agreement for $N=10$, as expected. However, at this stage, we begin to obtain original results: For $N=1$, high order quantum corrections are preponderant and the numerical results diverges from the analytical outcomes.
%\vspace{-0.5cm}
Furthermore, in the Fig.(\ref{Fig9pt1}) we also plot the variance of both the conductance and the shot-noise power. 
\begin{eqnarray}
\frac {var[G]}{G_0^2}&=&\frac{1}{8\beta}\left[1+\left(1-\Gamma\right)^2\right]
\end{eqnarray}
%\vspace{-1cm}
\subsection{Probability Distribution}
%\vspace{-0.5cm}

\begin{figure}[H]
\includegraphics[width=0.45\textwidth]{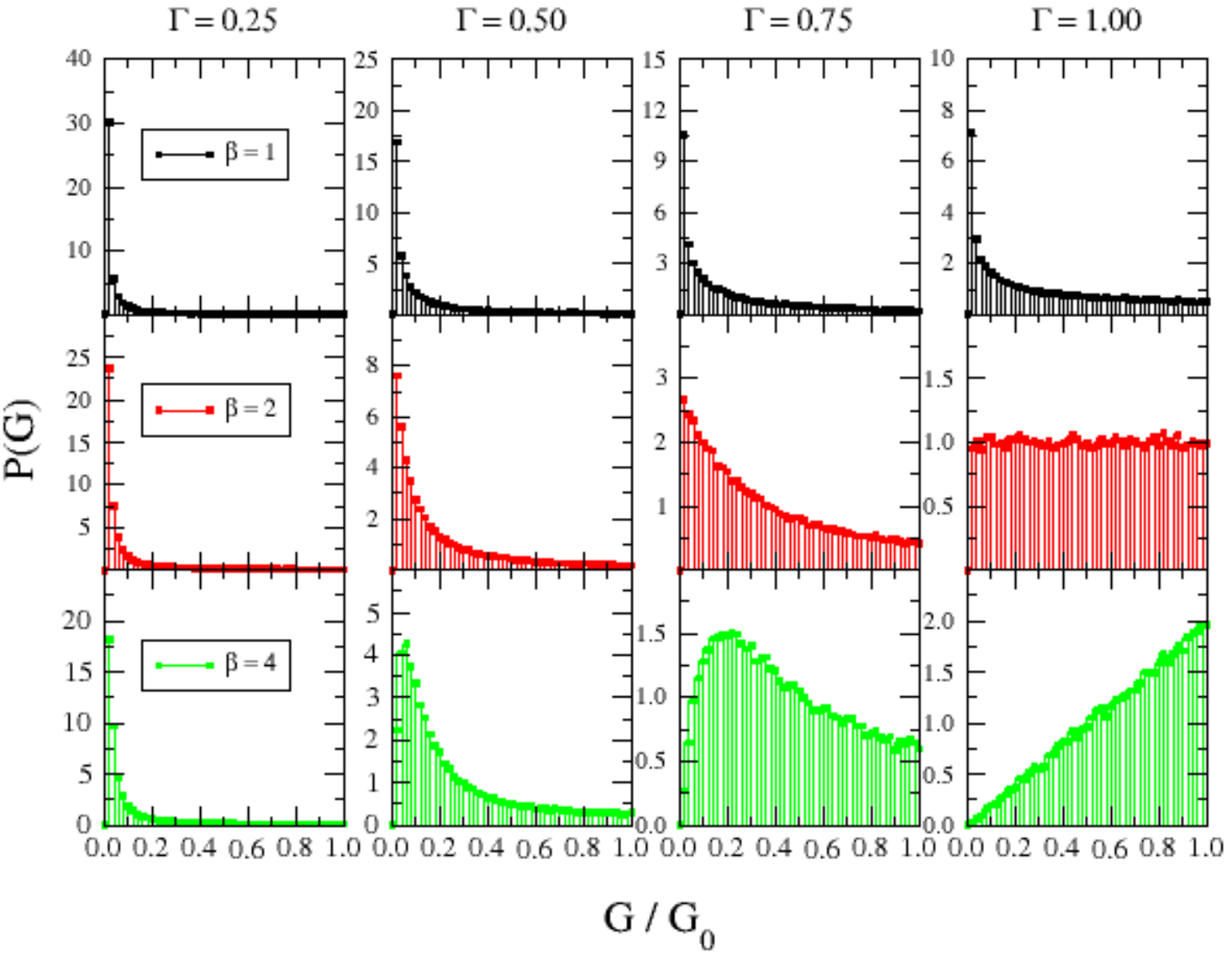}
\caption{Probability conductance distribution of the chaotic Schr\"odinger billiard for $\Gamma=0.25,0.5,0.75$ and $1.0$ in the symmetric leads regime $N_1=N_2=1$.} \label{Fig10}
\end{figure}

\begin{figure}[H]
\includegraphics[width=0.45\textwidth]{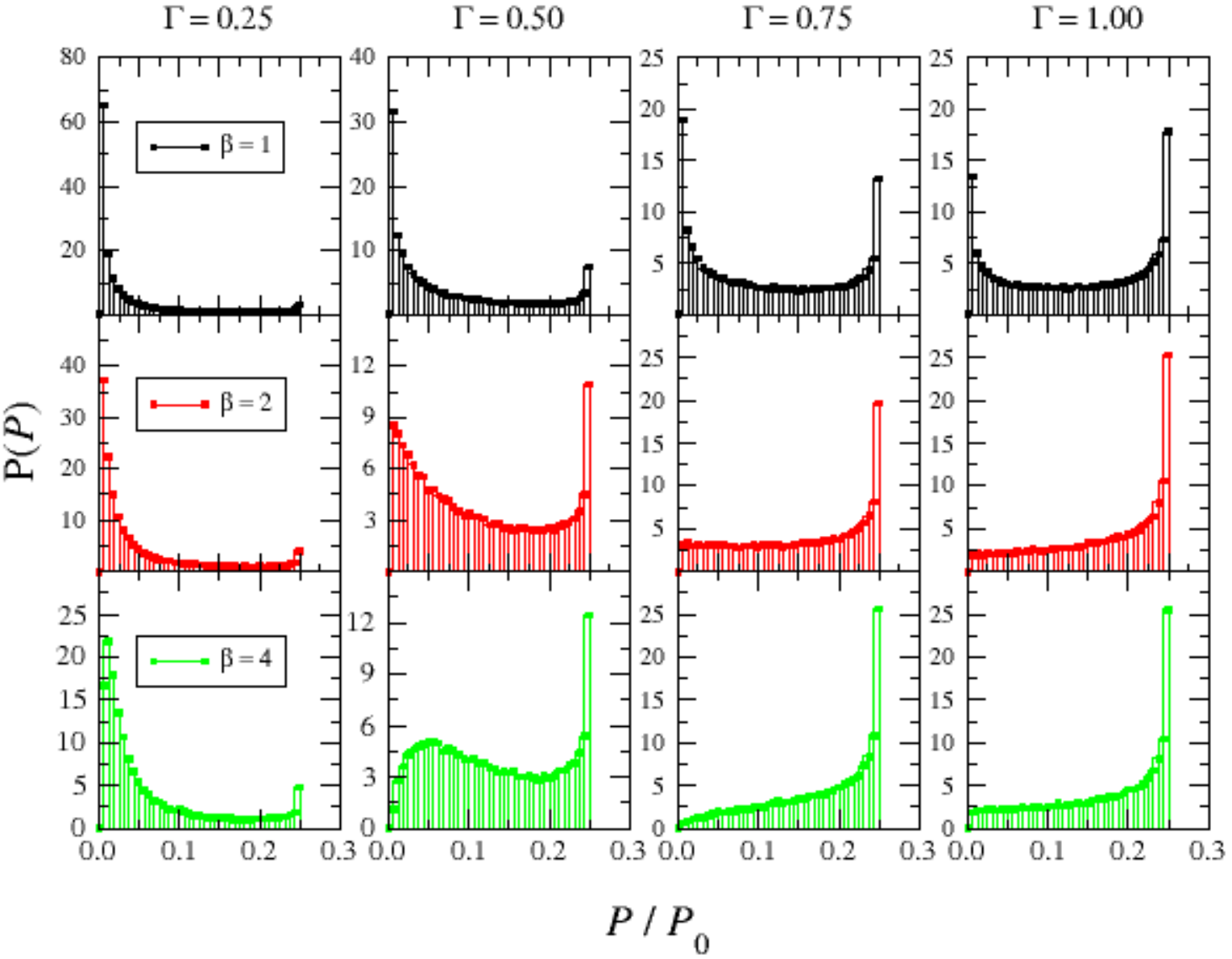}
\caption{Probability shot-noise distribution of chaotic Schr\"odinger billiard for  $\Gamma=0.25,0.5,0.75$ and $1.0$ in the symmetric leads regime $N_1=N_2=1$.} \label{Fig11}
\end{figure}

As demonstrated in the previous section, for few numbers of open channels in the leads, the electronic transport has a divergence with the well-known analytical results. Hence, the numerical investigation of the probability  distribution of both the conductance and the shot-noise in this regime gain a relevant role. Each realisation of the chaotic QD corresponds to a distinct value for all observables. Each of the values is generically known as the eigenvalue of the sample associated with the respective observable. In Figs.(\ref{Fig10}) and (\ref{Fig11}), we plot the conductance and shot-noise power distributions for $N=1$ and $\Gamma=0.25,0.5,0.75$ and $1.0$. 

As depicted in the Fig.(\ref{Fig10}), the modification of the tunneling barriers does not generates surprises concerning to the eigenvalues distribution of conductance: smaller barrier values (less transmission probability) tend to locate the eigenvalues close to zero, while the ideal case disperses the eigenvalues up to $G_0$ especially in $\beta=4$. The equivalent happens with the eigenvalues of the shot-noise power, as depicted in the Fig.(\ref{Fig11}), with the exception of the shot-noise larger eigenvalues which is always strongly spread.

\section{Non-Ideal Dirac Billiards} 

In this section, we show the effects of non-ideal contacts on the statistics of electronic transmission through a chaotic Dirac billiard.

\subsection{Conductance and Shot-Noise }

\begin{figure}[H]
\includegraphics[width=0.45\textwidth]{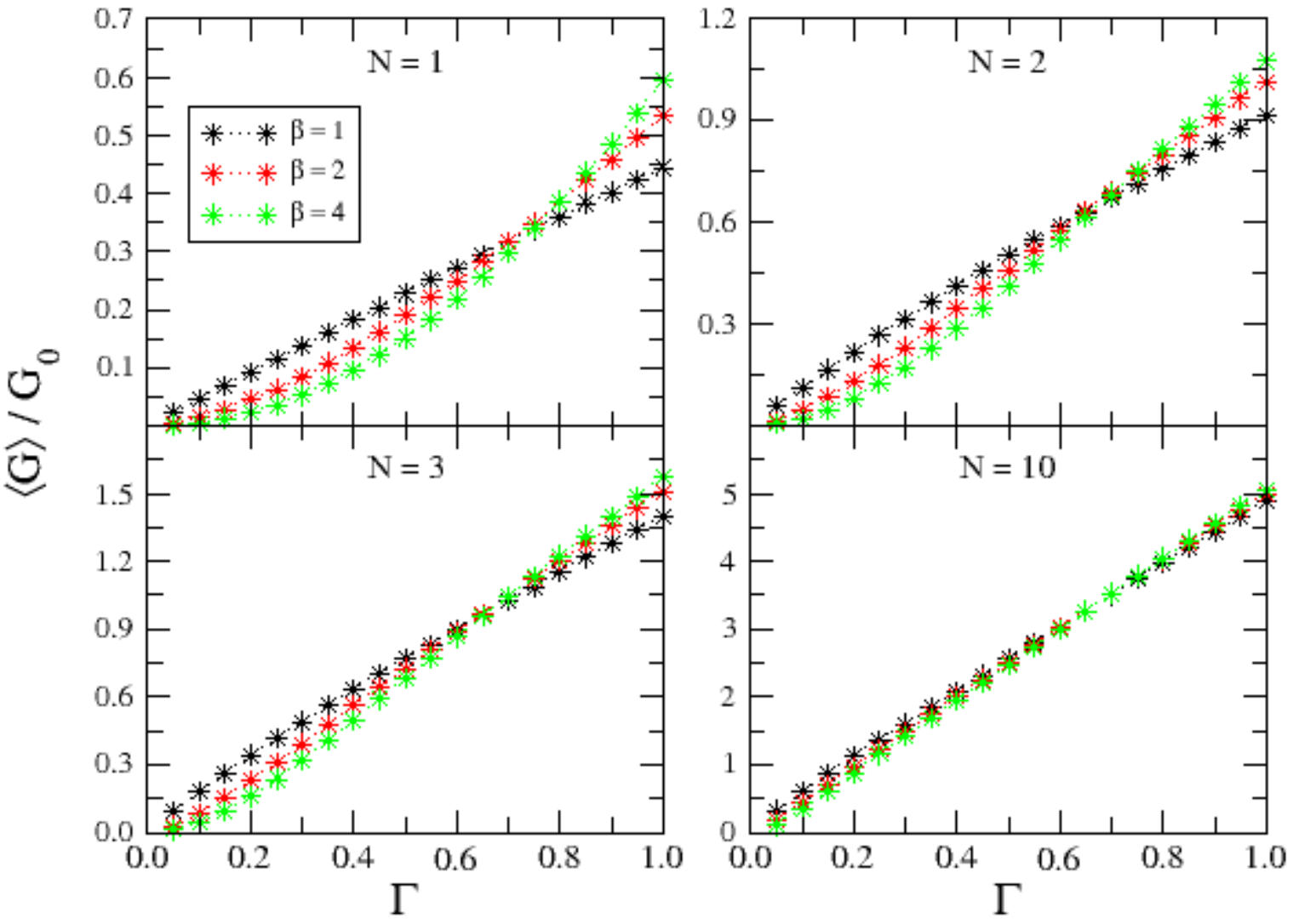}
\caption{The average of conductance as a function of tunnel barrier, $\Gamma$, for the Dirac billiard with symmetric leads $N=1,2,3$ and 10. The symbols are the numerical data obtained by random matrix theory.} \label{Fig12}
\end{figure}

\begin{figure}[H]
\includegraphics[width=0.45\textwidth]{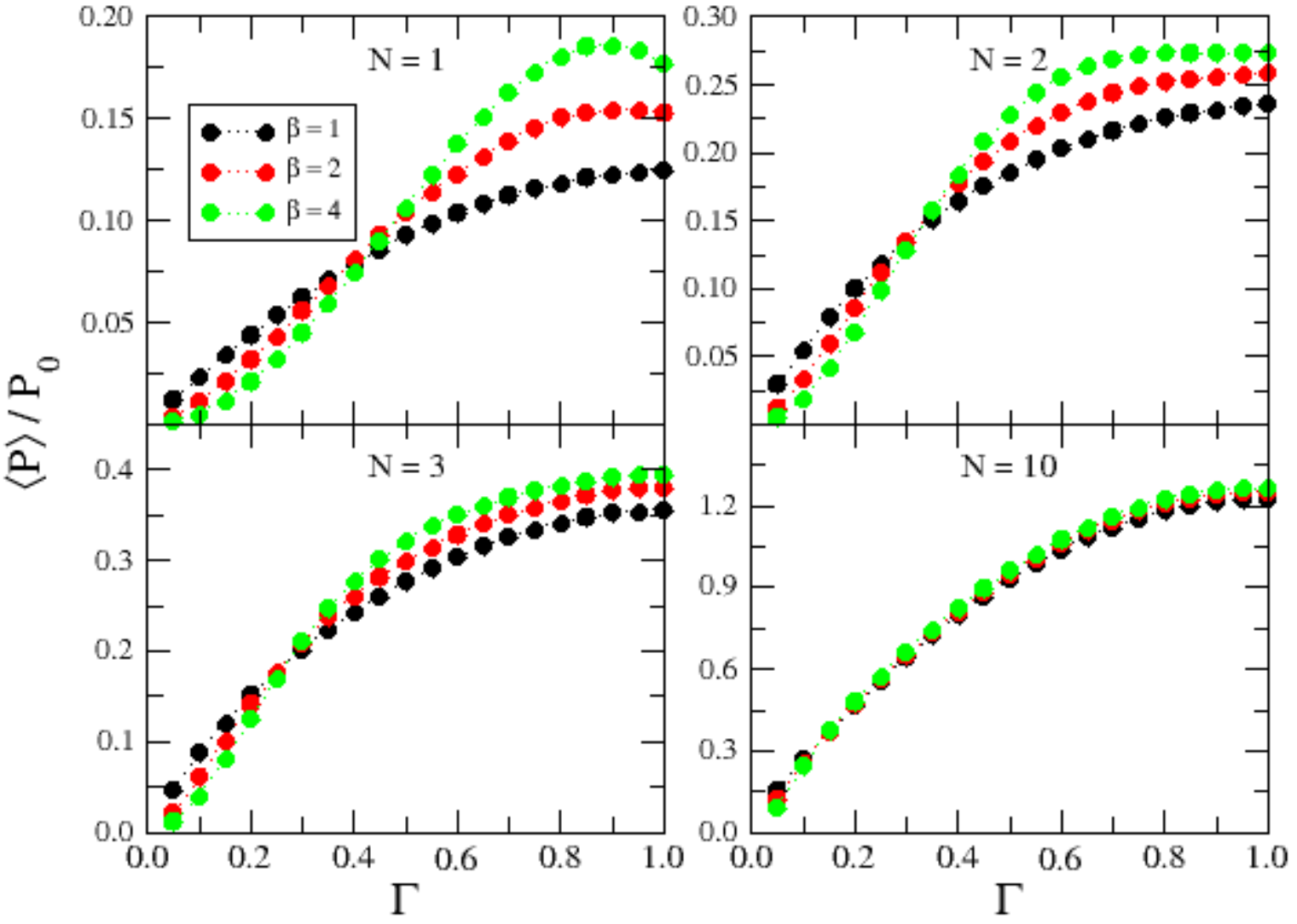}
\caption{The average of shot-noise power as a function of tunnel barrier $\Gamma$ for Dirac billiard with symmetric leads $N=1,2,3$ and 10. The symbols are numerical calculation data.} \label{Fig13}
\end{figure}

Despite many attempts, the analytical results on the Dirac billiard is, in general, limited to ideal contacts specially on the regime of few open channels, a non-integrable regime. For instance, the results to the average of conductance and the shot-noise power as a function of tunnel barriers are unknown until the moment. Therefore, there is a strong motivation for the numerical investigations and the obtention of general results in these systems, including graphene billiards in the non-ideal regime. In this section, we obtain, for the first time, the phenomenology of graphene billiard (Dirac billiard) in the non-ideal regime. 
Firstly, a prominent distinction is imperative, the quantum correction mechanism exhibited through the difference in the behaviour for few and large number of open channels. In the Figs.(\ref{Fig12}) and (\ref{Fig13}), we depict the results of our previously mentioned numerical simulation to the average of conductance and the shot-noise power as a function of the tunnel barriers for the three universal symmetries. Notice the difference between the curves both for de conductance and the shot-noise power as the regime of few open channels is reached. The difference is a signal of the preponderant role of the quantum interference corrections to the semi-classical therms as the width of the leads is reduced.

\begin{figure}[H]
\includegraphics[width=0.45\textwidth]{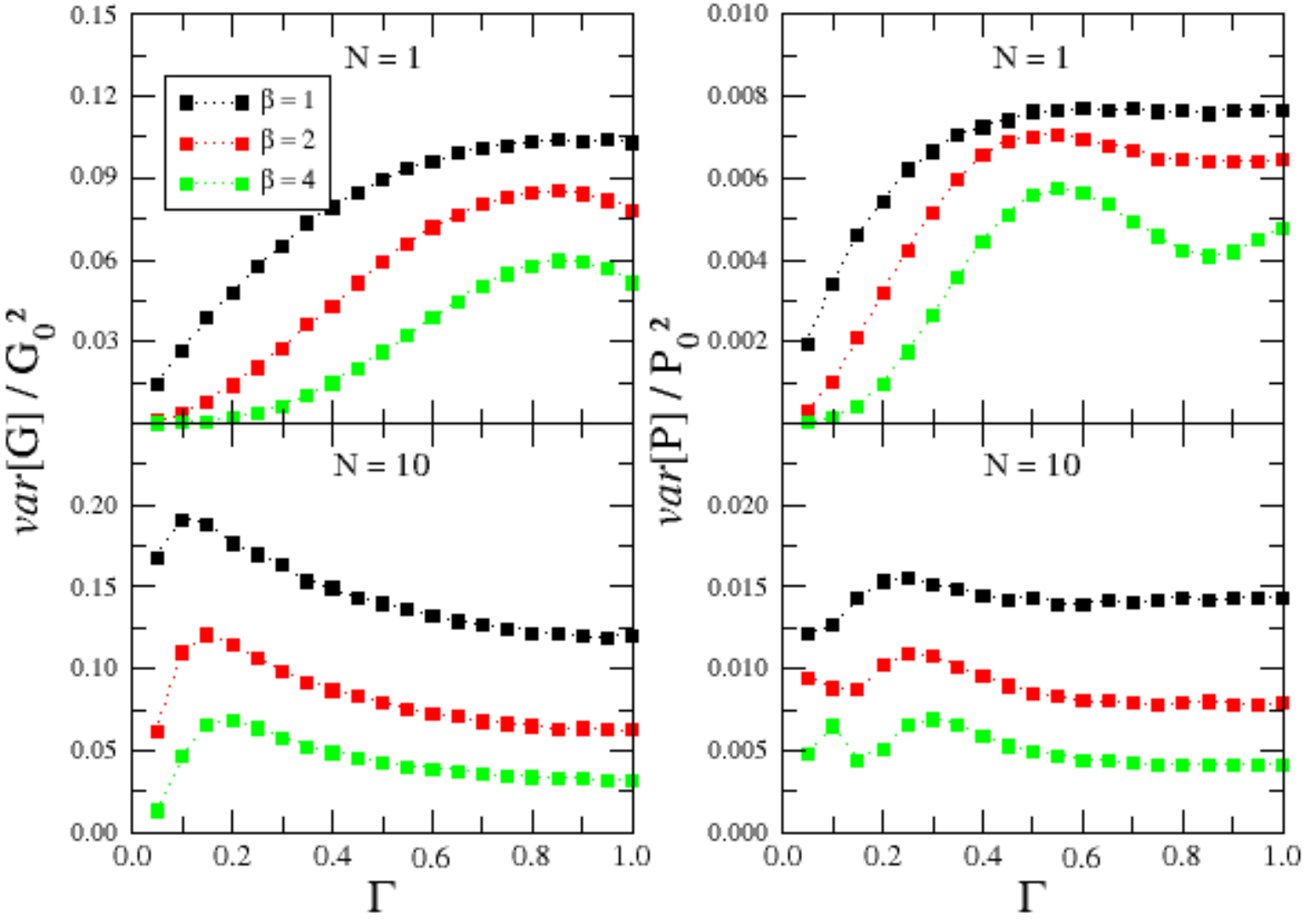}
\caption{The variance of conductance and shot-noise power for chaotic Dirac billiard.} \label{Fig14}
\end{figure}

Furthermore, we show a divergence in the Fig.(\ref{Fig14}) in which we plot the variance of conductance and shot-noise power. 

The Fig.(\ref{Fig12}) shows a counterintuitive property that has already been explored in the context of cosmology. The Klein paradox [\onlinecite{klein}] refers to the amplification of the tunnelling probability in relativistic systems if the potential barrier is increased. Notice that lower $\Gamma$ indicates a higher barrier value and, therefore, the suppression of both the classical and quantum terms of conductance or, correspondingly, indicates the decreasing of transport coefficients. However, as one observes in this figure, a striking effect occurs: For the same value of the potential barrier, the effect of the magnetic field is to amplify the conductance and not to produce a weak localization (a result expected from the Anderson theory of localization), as we can see in the Fig.(\ref{Fig12}). This indicates that the quantum interference correction term (QICT) is highly anomalous in the Dirac billiard.

\begin{figure}[H]
\includegraphics[width=0.45\textwidth]{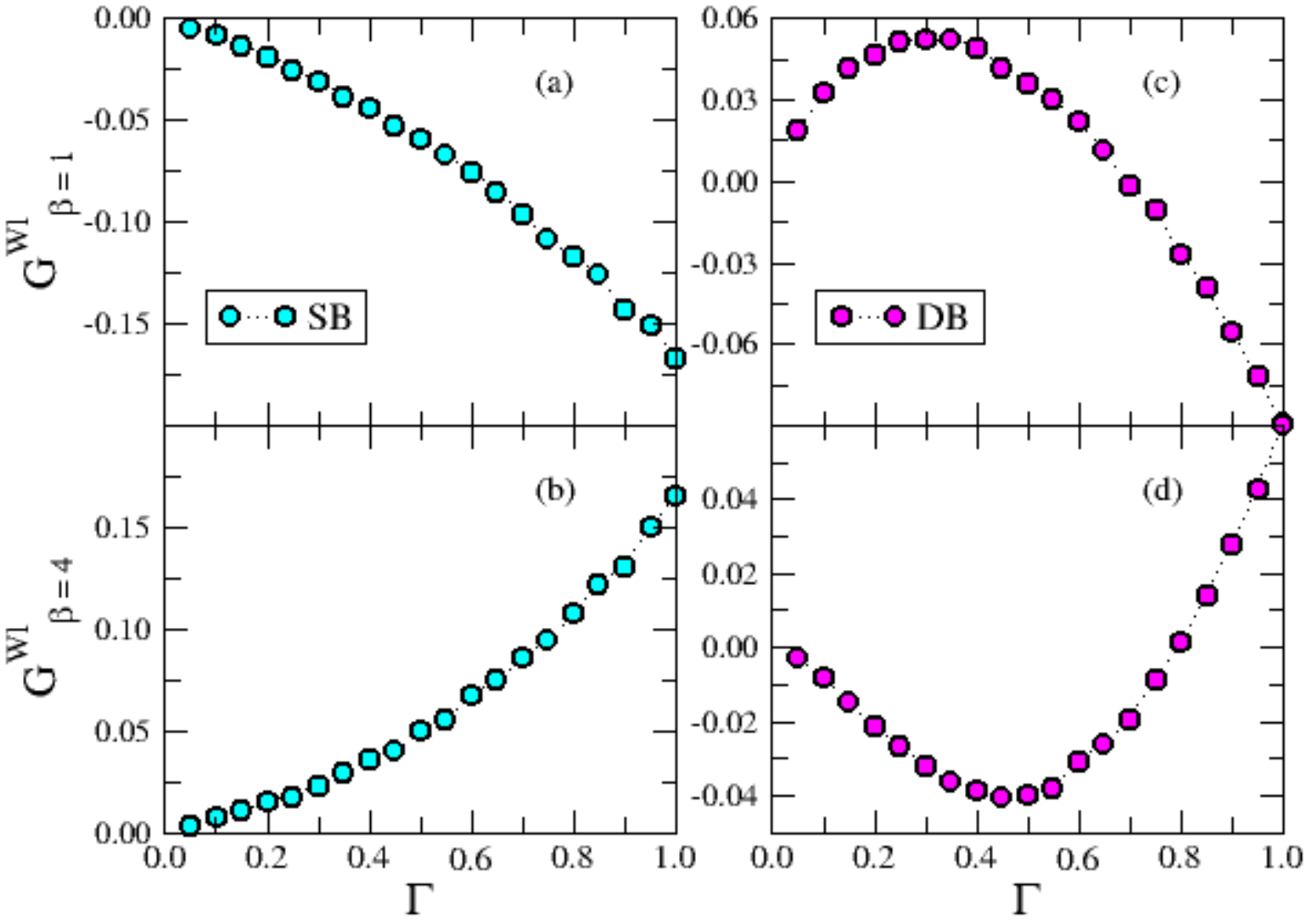}
\caption{Localization of conductance for Schr\"odinger billiard ($a$ and $b$) and Dirac billiard ($c$ and $d$).} 
\label{Fig18Gpt0}
\end{figure}

We obtain the QICT subtracting the averaged values ​​obtained in any ensemble from the ones obtained in $\beta=2$ ensemble which contains only the classical term. Observing the Fig.(\ref{Fig18Gpt0}), we find that the increase of the barrier produces an amplification of the quantum interference term. This indicates a very peculiar type of Klein paradox exhibited in Dirac billiard and not observed in Schr\"odinger billiard. Therefore, the chaotic Klein paradox can be measured by the inversion of quantum anti-localization caused in a very special way by the time-reversal symmetry breaking.

\begin{figure}[H]
\includegraphics[width=0.45\textwidth]{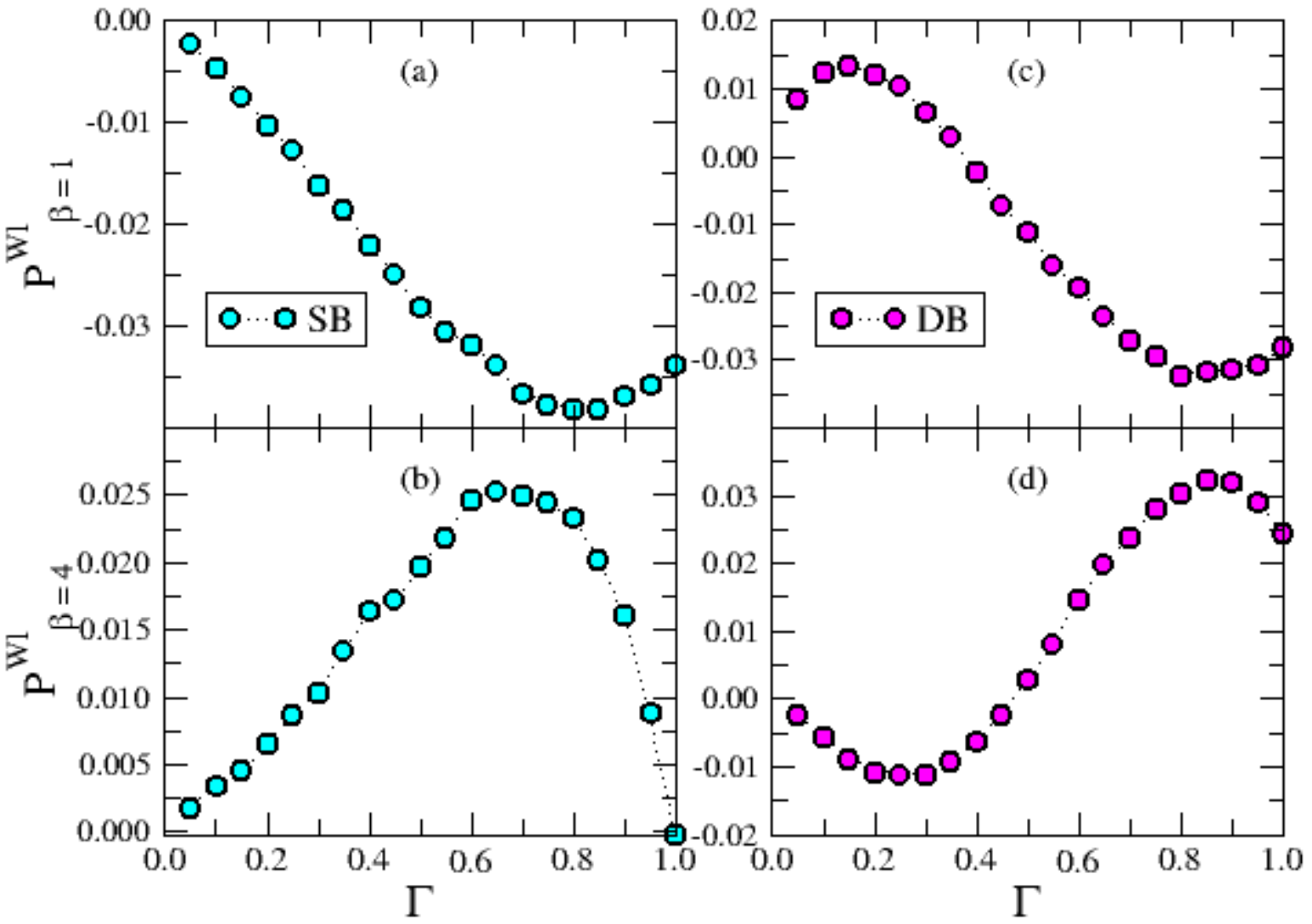}
\caption{Localization of shot-noise power for Schr\"odinger billiard ($a$ and $b$) and Dirac billiard ($c$ and $d$).} \label{Fig18Ppt0}
\end{figure}

In the reference [\onlinecite{sugestao1}], the authors showed interesting electronic propagation paths induced by the Klein paradox in graphene point contact subject to external potentials. In particular, these abnormal paths may lead to conductance fluctuations with fractal-like behaviour that do not appear in usual two-dimensional structures. Accordingly, the electrons tend to traverse regions with greater potential (edges of the constriction), unlike what happens in Schr\"odinger's structures. In the reference [\onlinecite{sugestao2}], the authors solve the resonant two-dimensional Dirac equation for two symmetric quantum dots separated by a tunneling barrier. They show that chaos can suppress the effect of quantum tunneling, leading to a regularization of the tunneling dynamics including graphene. These effects, jointly with our results in klein paradox in the presence of universal chaos, suggest that there is an emerging competition between chaos, tunneling, confinement, and the underlying symmetries that determine the essence of quantum transport.

\subsection{Probability Distribution}

\begin{figure}[H]
\includegraphics[width=0.45\textwidth]{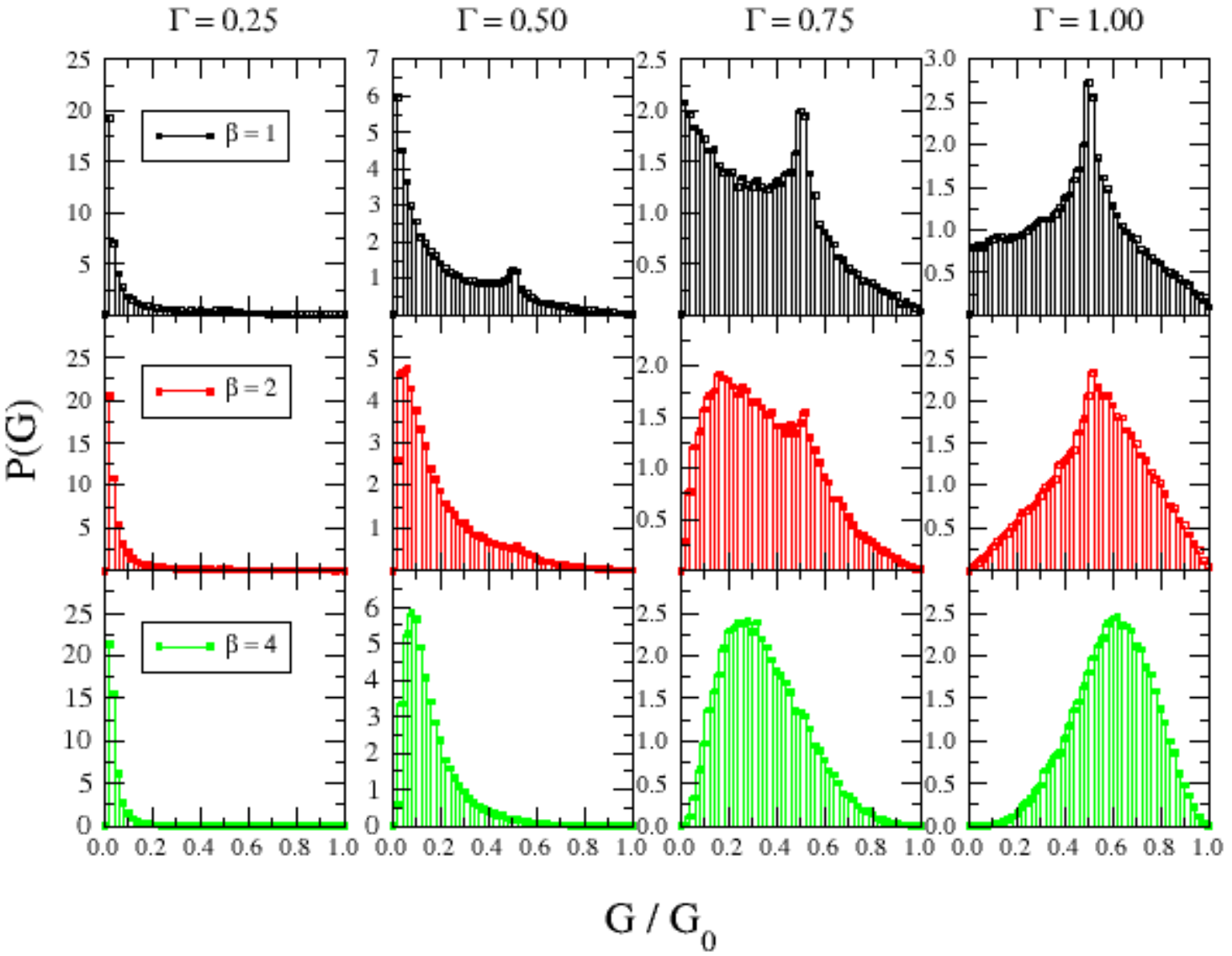}
\caption{Probability conductance distribution of chaotic Dirac billiard for  $\Gamma=0.25,0.5,0.75$ and $1.0$ in the symmetric leads regime $N_1=N_2=1$.} \label{Fig15}
\end{figure}

\begin{figure}[H]
\includegraphics[width=0.45\textwidth]{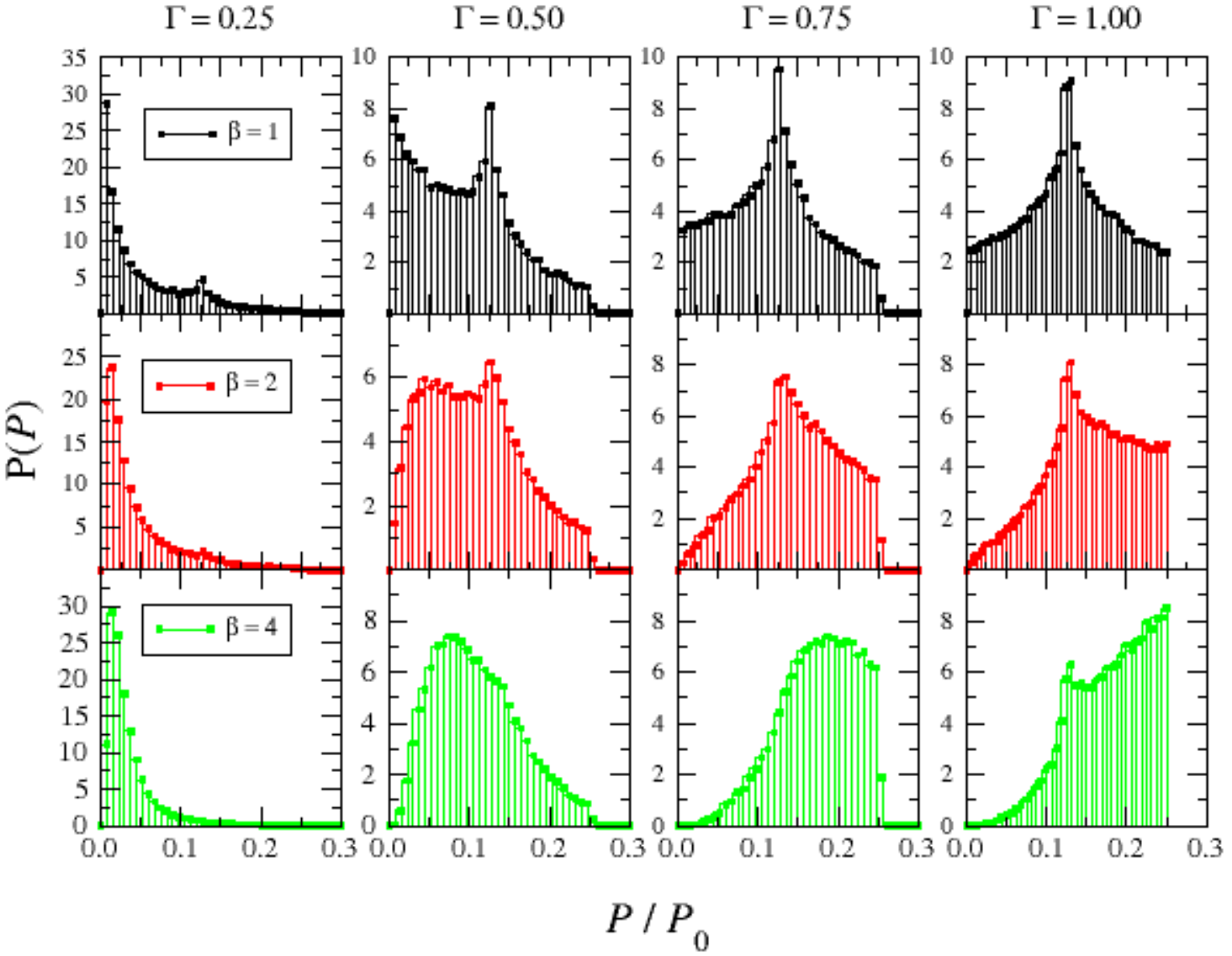}
\caption{Probability shot-noise distribution of chaotic Dirac billiard for  $\Gamma=0.25,0.5,0.75$ and $1.0$ in the symmetric leads regime $N_1=N_2=1$.} \label{Fig16}
\end{figure}

As demonstrated in the previous section, for few numbers of open channels in the leads the electronic transport has a very peculiar behavior. Hence, the numerical study of probability distribution of conductance and shot-noise in this regime is again important. In the Figs.(\ref{Fig15}) and (\ref{Fig16}), we plot the conductance and shot-noise power distributions for $N=1$ and $\Gamma=0.25,0.5,0.75,1.0$. 

According to Fig.(\ref{Fig15}), the distribution of conductance eigenvalues of a Dirac billiard tends to locate around zero for small values of $\Gamma$ (low probability of transmission), as expect. As the $\Gamma$ value increases, the eigenvalues disperse. However, surprisingly, the eigenvalues tend to appear in $G \approx 0.5 G_0$ and not in the maximum, $G=G_0$, as expected for the ideal systems and confirmed in the Schr\"odinger billiard. In addition, around $G \approx 0.5 G_{0}$ appears an anomalous singularity (not present in any symmetries of WD ensembles) for $\beta=1$ and $\beta=2$. Therefore, structures such as graphene determine anomalous singularities in the distribution of conductance eigenvalues. The Fig.(\ref{Fig16}) shows the shot-noise power distribution eigenvalues, which eliminates the peak in the maximum value (present in WD symmetries) and creates a strong peak in the intermediate values in which, like in Dirac billiard conductance, an anomalous singularity in both $\beta=1$ and in $\beta=2$ is created.

%\vspace{-11cm}
\section{Conclusion}
%\vspace{-2cm}
In this paper, we performed a complete and self-contained numerical investigation of both the chaotic Dirac Billiards and the Schr\"odinger Billiards. By direct comparisons between the two kinds of scattering, we observed anomalous behaviors in the conductance, the shot-noise power, the universal conductance fluctuations and the respective eigenvalues distributions. We especially observed a myriad of relevant singularities and anomalous concentrations in the eigenvalues distributions in the Dirac billiards, inversions and suppression/amplification transitions in the conductance and in the shot-noise power. More importantly, we identified the Klein paradox that can be measured and controlled in experiments involving graphene quantum dots. We hope that our results will have many unfolding in the endeavour into the understanding of phenomena involving universal resonant quantum scattering and chaos in the mesoscopic regime in both non-relativistic and relativistic systems.

This work was partially supported by CNPq, CAPES and FACEPE (Brazilian Agencies).

\end{document}